\documentclass[twocolumn,showpacs,preprintnumbers,prl]{revtex4}

\usepackage{graphicx}
\usepackage{amsmath}
\usepackage{amssymb}
\usepackage{natbib}

\begin{document}

\title{Vibrations and diverging length scales near the unjamming
  transition}

\author{Leonardo E. Silbert}\email{lsilbert@uchicago.edu}

\affiliation{James Franck Institute, University of Chicago,
  Chicago, IL 60637}

\affiliation{Department of Chemistry and Biochemistry, UCLA, Los
  Angeles, CA 90095}

\author{Andrea J. Liu}

\affiliation{Department of Physics and Astronomy, University of
  Pennsylvania, Philadelphia, PA 19104}

\author{Sidney R. Nagel}

\affiliation{James Franck Institute, The University of Chicago,
  Chicago, IL 60637}

\date{\today}

\begin{abstract}
  
  We numerically study the vibrations of jammed packings of particles
  interacting with finite-range, repulsive potentials at zero
  temperature. As the packing fraction $\phi$ is lowered towards the
  onset of unjamming at $\phi_{c}$, the density of vibrational states
  approaches a non-zero value in the limit of zero frequency.  For
  $\phi>\phi_{c}$, there is a crossover frequency, $\omega^{*}$ below
  which the density of states drops towards zero. This crossover
  frequency obeys power-law scaling with $\phi-\phi_{c}$.
  Characteristic length scales, determined from the dominant
  wavevector contributing to the eigenmode at $\omega^{*}$, diverge as
  power-laws at the unjamming transition.
  
\end{abstract}


\maketitle

The jamming/unjamming transition for zero-temperature sphere packings
has a mixed first-order/second-order character \cite{ohern2,ohern3}.
This transition occurs for particles with finite-range and purely
repulsive interactions. As a system unjams with decreasing packing
fraction, the number of interacting neighbors per particle drops
discontinuously to zero. Despite this characteristic signature of
first-order behavior, power-law scaling is also observed for other
quantities \cite{ohern2}. This raises the question of whether there is
a diverging length scale associated with the loss of rigidity.
Simulations suggest that a diverging length scale exists on the
low-density side of the transition \cite{ohern3,reichhardt1}, but
there has been no demonstration of similar behavior in the jammed
phase. Because the jamming transition may control the glass transition
at higher densities and temperatures \cite{ohern3}, an observation of
a diverging correlation length should shed light on the nature of the
glass transition and properties of glasses in general. Here we show
that a diverging length can be extracted from the vibrational
properties of the jammed phase.

In most crystalline or amorphous solids, vibrations at low frequency,
$\omega$, are expected to be long-wavelength, acoustic plane waves.
From this assumption one obtains the asymptotic low-frequency Debye
density of vibrational states: $\mathcal{D}(\omega) \propto
\omega^{D-1}$ where $D$ is the dimension of space. A feature of glassy
systems is that there is an excess in the density of low-frequency
vibrations \cite{phillips1}. There has been an extensive body of
simulation work on phonon spectra in quenched glasses dating back to
the 1970's \cite{mctague1}. In particular, studies of the
Lennard-Jones glass found an increasing number of low-frequency modes
as the system was diluted in an {\it ad-hoc} fashion \cite{nagel8}.
Perhaps the most dramatic demonstration of this excess density of
low-frequency modes appears in a system at zero temperature with a
well-defined jamming transition at packing fraction $\phi_{c}$
\cite{ohern3}. For systems with attractions, the jamming transition is
usually inaccessible because of the vapor-liquid phase
transition~\cite{sastry1}.  Therefore, a system with soft,
purely-repulsive, finite-range interactions has the advantage of
allowing a systematic study of the properties near its well-defined
jamming transition.

In the simulations reported here we have studied monodisperse spheres
of diameter $\sigma$ interacting through a finite-range, purely
repulsive, harmonic potential:
\[
V(r)= \{
\begin{array}{lrr}
  V_{0}(1- r/\sigma)^{2} && r < \sigma \\
  0 && r \geq \sigma.
\end{array}
\]
   
Here $r$ is the center-to-center separation between two particles.
The units of length and time are $\sigma$ and $(md^{2}/V_{0})^{1/2}$,
respectively, for particles of mass $m$.  Our three dimensional (3D)
systems consist of $N =$ 1024, 4096, and 10000 spheres in cubic
simulation cells with periodic boundary conditions. We employ
conjugate-gradient energy minimization in order to obtain $T = 0$
configurations.  We have also studied bidisperse, harmonic discs in
$2D$, as well as Hertzian contact potentials in $3D$.  Our conclusions
are the same for all three systems.

\begin{figure}[h]
  \includegraphics[width=6.5cm]{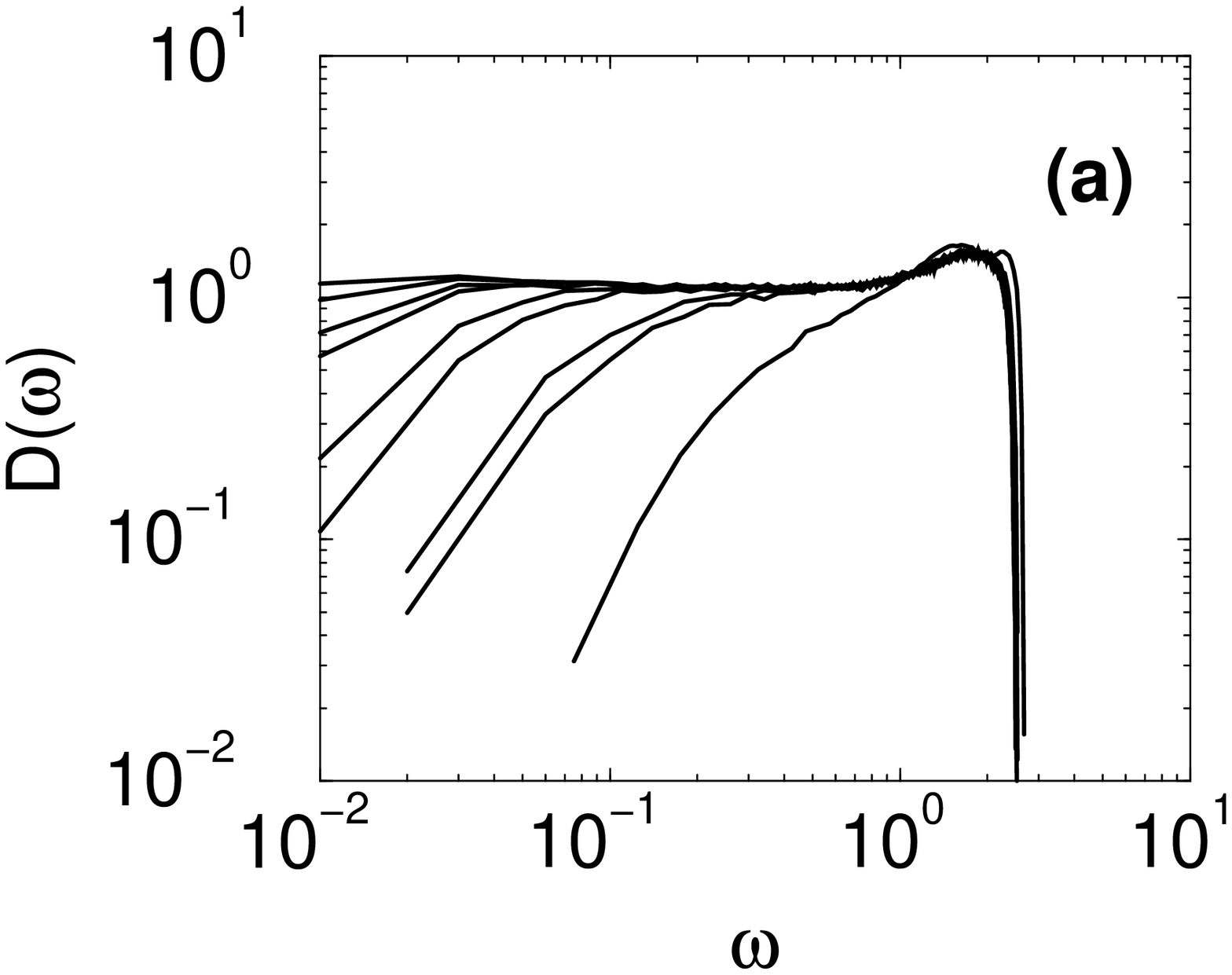}
  \includegraphics[width=6.5cm]{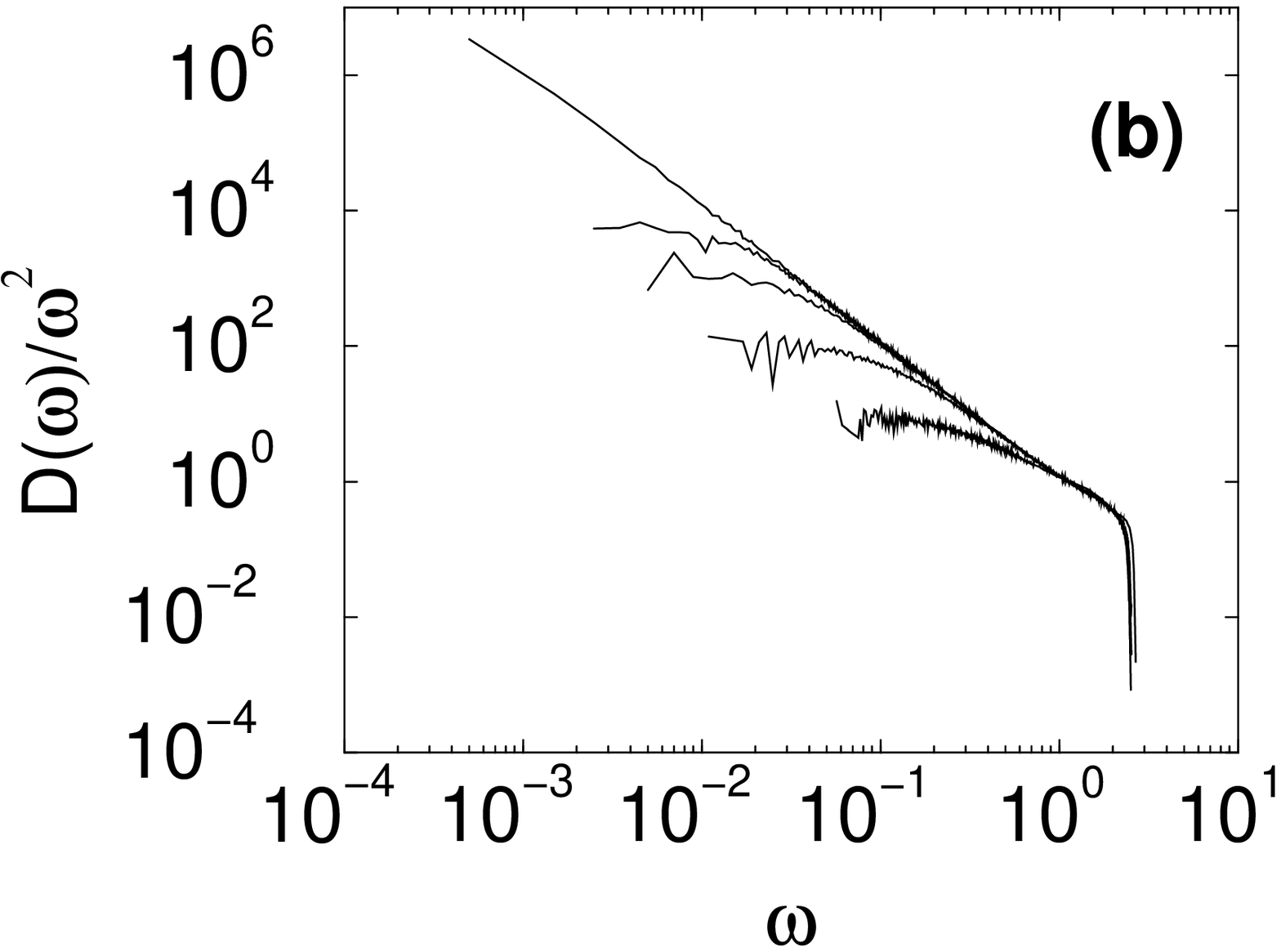}
  \caption{(a) Density of vibrational states $\mathcal{D}(\omega)$, for
    $N=1024$ monodisperse spheres in $3D$, interacting via harmonic
    repulsions. The curves from right to left are for, $\phi -
    \phi_{c} = 1\times 10^{-1},~1\times 10^{-2},~5\times
    10^{-3},~1\times 10^{-3},~5\times 10^{-4},~1\times
    10^{-4},~5\times 10^{-5},~1\times 10^{-6},~1\times 10^{-8}$. (b)
    $\mathcal{D}(\omega)/\omega^{2}$ at $\phi - \phi_{c} = 1\times
    10^{-6},~ 1\times 10^{-4},~ 1\times 10^{-3},~ 1\times 10^{-2}$,
    and $1\times 10^{-1}$, using data from $N = 1024,
    4096,~\text{and}~10000$.}
  \label{fig1}
\end{figure}

In Fig.~\ref{fig1}(a) we show a log-log plot of the density of states,
$\mathcal{D}(\omega)$ as a function of $\omega$ covering eight decades
in $(\phi - \phi_{c})$ for one system size. For the smallest value of
$(\phi - \phi_{c})$ studied, the low-frequency behavior approaches a
flat spectrum with an $\omega = 0$ intercept of $\mathcal{D}_{0}
\equiv \mathcal{D}(\omega \rightarrow 0)$. Thus, close to the jamming
transition there is no longer any vestige of Debye behavior. As the
system is compressed above threshold, the curves depart from this
plateau behavior at a frequency $\omega^{*}$, which increases with
$(\phi - \phi_{c})$. Below $\omega^{*}$,
$\mathcal{D}(\omega)\rightarrow 0$ as $\omega \rightarrow 0$.
Experiments on vitreous silica \cite{buchenau4} and simulations of
models of glasses with three-body interactions \cite{pilla1} have
observed a similar trend with decreasing pressure. Figure
\ref{fig1}(b) is the full spectrum divided by $\omega^{2}$ for all
system sizes, at several values of $(\phi - \phi_{c})$. The peak
position in $\mathcal{D}(\omega)/\omega^{2}$ shifts to lower
frequencies and the peak height increases as $\phi_{c}$ is approached,
analogous to the way some glasses behave as the glass transition
temperature is approached \cite{buchenau5}.  Thus, our
zero-temperature system captures features often associated with the
Boson peak in real glasses \cite{elliott1}.

\begin{figure}[h]
  \includegraphics[width=6.5cm]{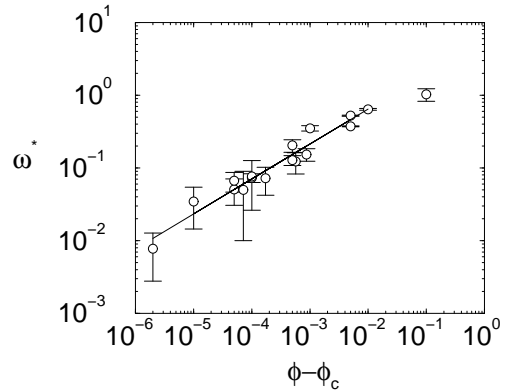}
  \caption{The crossover frequency $\omega^{*}$. The line is a fit to
    Eq.~\ref{equation1} with $\Omega = 0.48$.}
  \label{fig2}
\end{figure}

Figure \ref{fig2} shows the crossover frequency, $\omega^{*}$, versus
$(\phi - \phi_{c})$. We determine $\omega^{*}$ by finding where
$\mathcal{D}(\omega)$ for a given $(\phi - \phi_{c})$ departs from
$\mathcal{D}(\omega)$ for the next smallest value of $(\phi -
\phi_{c})$. Over more than 4 decades in $(\phi - \phi_{c})$,
$\omega^{*}$ vanishes as a power law:
\begin{equation}
  \omega^{*} \propto (\phi - \phi_{c})^{\Omega}
  \label{equation1}
\end{equation}
with $\Omega= 0.48 \pm 0.03$.  The scaling is robust, independent of
the precise manner in which we determine $\omega^{*}$. It is the same
(with different prefactors) when $\omega^{*}$ is defined as the value
of $\omega$ at which $\mathcal{D}(\omega)$ reaches $D_{0}$, $0.9
D_{0}$ or $0.5 D_{0}$, as well as the frequency where
$\mathcal{D}(\omega) / \omega^{2}$ in Fig.~\ref{fig1}(b) levels off
with decreasing $\omega$.

\begin{figure}[h]
  \includegraphics[width=6.5cm]{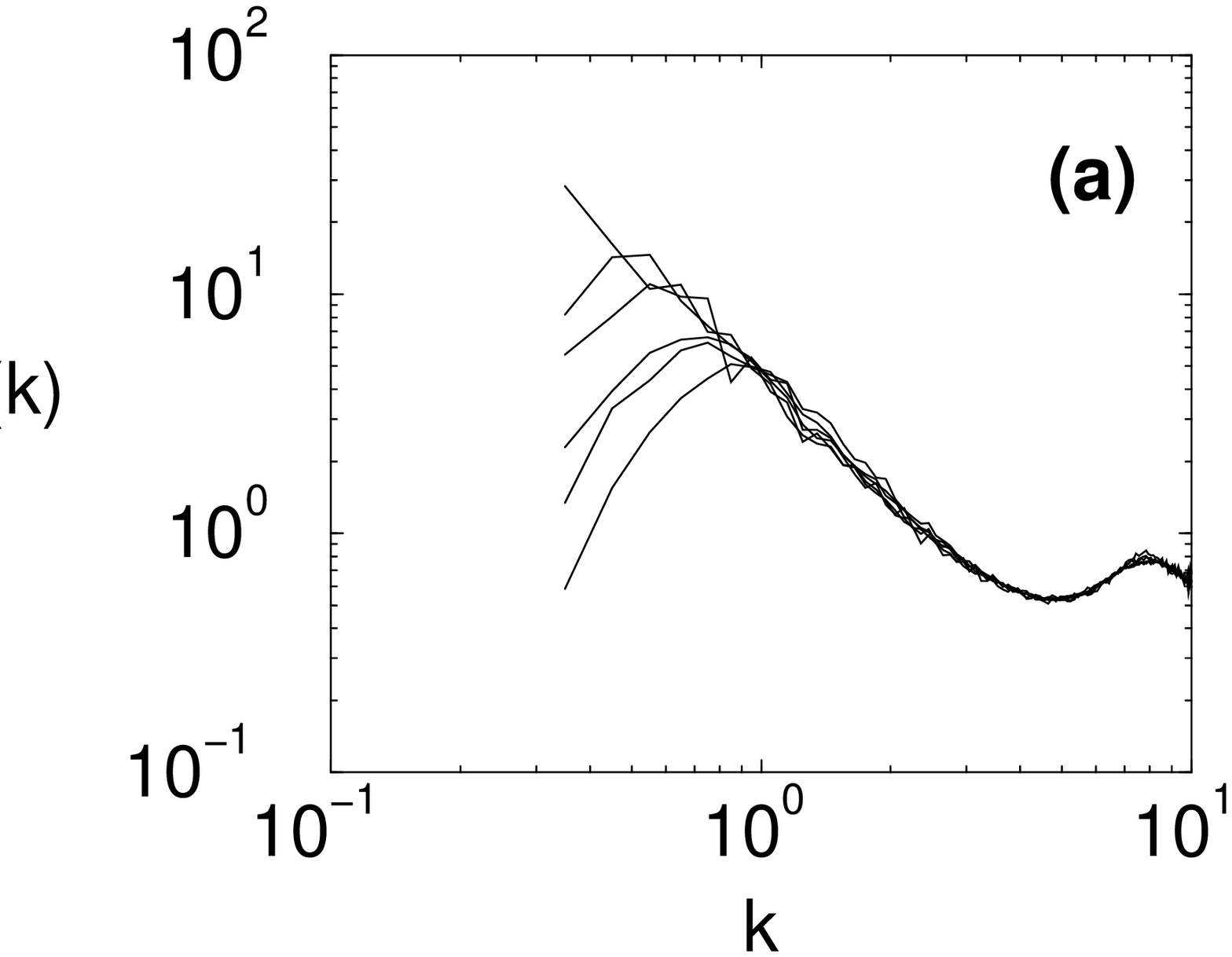}
  \includegraphics[width=6.5cm]{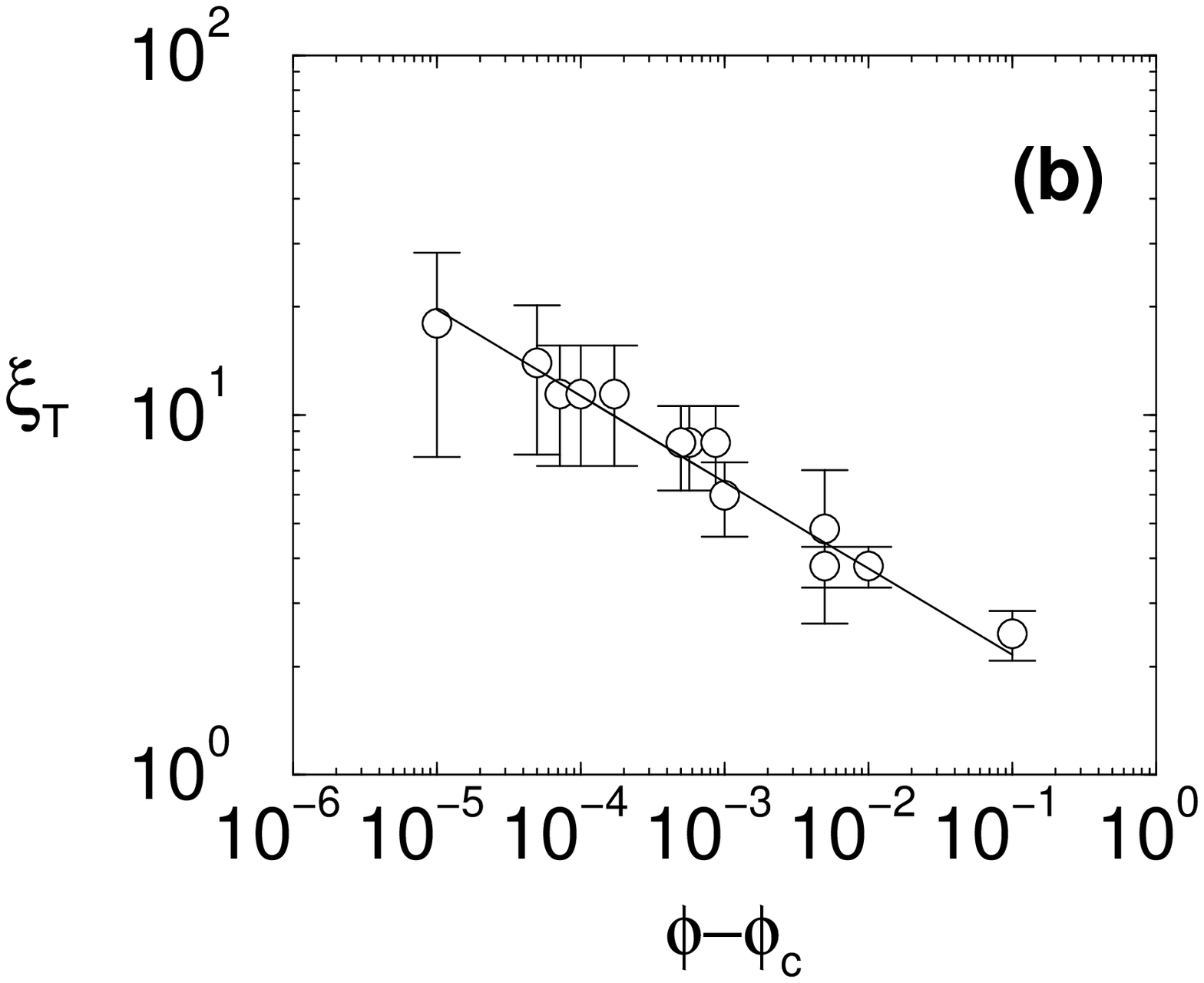}
  \caption{(a) Transverse spectral functions $f_{T}(k,\omega^{*})$ for
    $\phi - \phi_{c} = 1\times 10^{-5},~5\times
    10^{-5},~1\times 10^{-4},~5\times 10^{-4},~1\times 10^{-3},~5\times
    10^{-3}$, with decreasing amplitude respectively. (b) The
    correlation length $\xi_{T}\equiv 2\pi/k^{*}$ obtained from the
    wave-vector $k^{*}$ for the position of the peak in $f_{T}$ at
    frequency $\omega^{*}$.}
  \label{fig3}
\end{figure}

Given $\omega^{*}$, we can define a corresponding length scale,
$\xi_{T}$, that diverges as $\phi$ approaches $\phi_c^{+}$. In order
to extract $\xi_{T}$ we examine the spatial variation of the eigenmode
corresponding to the frequency $\omega^{*}$. We take the Fourier
transform of the appropriate component of the polarization vector,
${\bf P}_{i}(\omega^{*})$ of each particle $i$, which, for transverse
waves is,
\[
f_{T}(k,\omega^{*}) = \left\langle \left| \sum_{i} {\bf\hat{k}} \wedge
    {\bf P}_i(\omega^*) \exp(\imath {\bf k} \cdot {\bf r}_i)\right|^2
\right\rangle,
\]
where $\langle \; \rangle$ denotes an average over configurations and
over all wave-vectors with the same magnitude $k$. (The longitudinal
component, $f_{L}(k,\omega)$, not shown \cite{leo14}, is the dynamical
structure factor accessible from inelastic neutron scattering.)
Figure \ref{fig3}(a) shows these functions at the values of
$\omega^{*}$ determined for different compressions, $(\phi -
\phi_{c})$. The transverse components exhibit well-defined peaks at
$k^{*}$ at small wavevectors (there is a system size cut-off at
$k_{min}=2\pi/L$, where $L$ is the size of the simulation box). Thus,
$\xi_{T} \equiv 2 \pi/k^{*}$ is the dominant transverse length scale
for that mode.  Figure \ref{fig3}(b) shows $\xi_{T}$ as a function of
$(\phi - \phi_{c})$. The solid line corresponds to a power-law fit:
\begin{equation}
  \xi_{T} \propto (\phi - \phi_{c})^{-\nu_{T}}~~,
  \label{equation2}
\end{equation}
with $\nu_{T} = 0.24 \pm 0.03$. The exponents $\nu_{T}$ and $\Omega$
(Eq.~\ref{equation1}) can be related to each other via a simple
scaling argument using $k^{*}=\omega^{*}/c_{\text{T}}(\phi)$, where
$c_{\text{T}}(\phi)$ is the velocity of transverse sound. The shear
modulus $G_{\infty}$ vanishes with an exponent of $\gamma=0.48 \pm
0.05$ \cite{ohern3}, implying that $c_{\text{T}}(\phi)$ vanishes with
an exponent of 0.24. Given $\Omega=0.48$, we would therefore expect
$\nu_{T}=\Omega-\gamma/2 \approx 0.24$, in agreement with the results
of Fig.~\ref{fig3}(b).

A similar argument can be constructed for a longitudinal length
$\xi_{L}$ based on the peak of $f_{L}(k,\omega^{*})$. Although it is
difficult to extract the peak value from $f_{L}(k,\omega^{*})$ because
the peaks occur at very low values of $k$, the analogous scaling
relation based on the longitudinal sound speed (and hence the bulk
modulus, which is independent of $\phi-\phi_{c}$ \cite{ohern3})
predicts
\begin{equation}
  \nu_{L} \approx 0.48.
  \label{equation3}
\end{equation}

We have also simulated a system with Hertzian potentials: $V_{0}(1-
r/\sigma)^{5/2}$, for $r<\sigma$, and zero otherwise. In this case, we
find that $\omega^{*}$ vanishes with an exponent $\Omega \approx 0.78
\pm 0.03$, which is different from that obtained for the harmonic
case. However, by calculating the peak of $f_{T}(k,\omega^{*})$ we
obtain a length scale $\xi_{T}$ that diverges with $\nu_{T} \approx
0.23$, just as in the harmonic case. This is again consistent with the
scaling argument based on the speed of sound, since the shear modulus
vanishes with an exponent $\gamma\approx 0.95$ for the Hertzian case
\cite{ohern3}. Thus, even though the scaling of $\omega^{*}$ is
different for the Hertzian and harmonic potentials, we have the same
scaling for $\xi_{T}$.

\begin{figure}[h]
  \includegraphics[width=4cm]{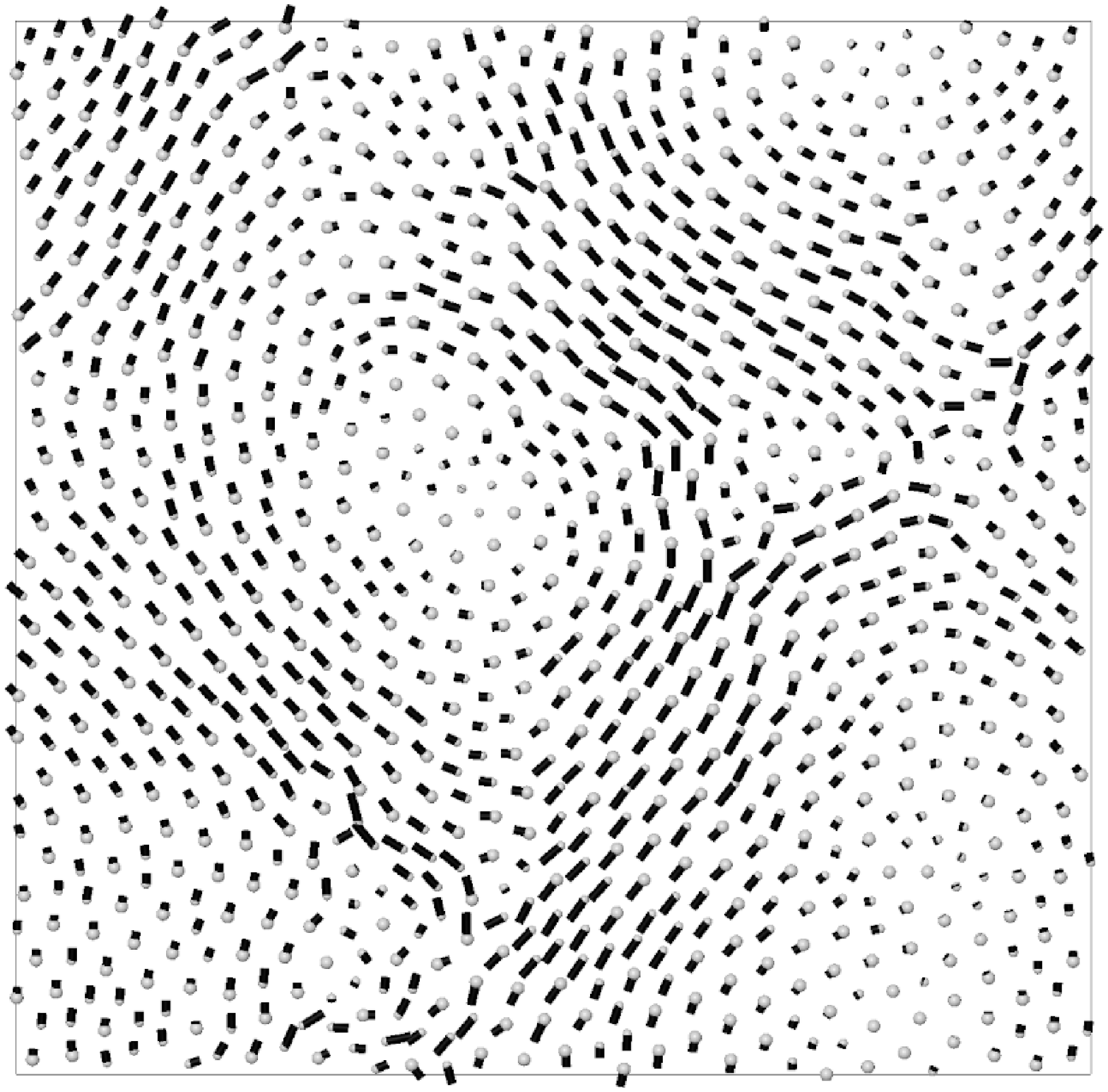}
  \includegraphics[width=4cm]{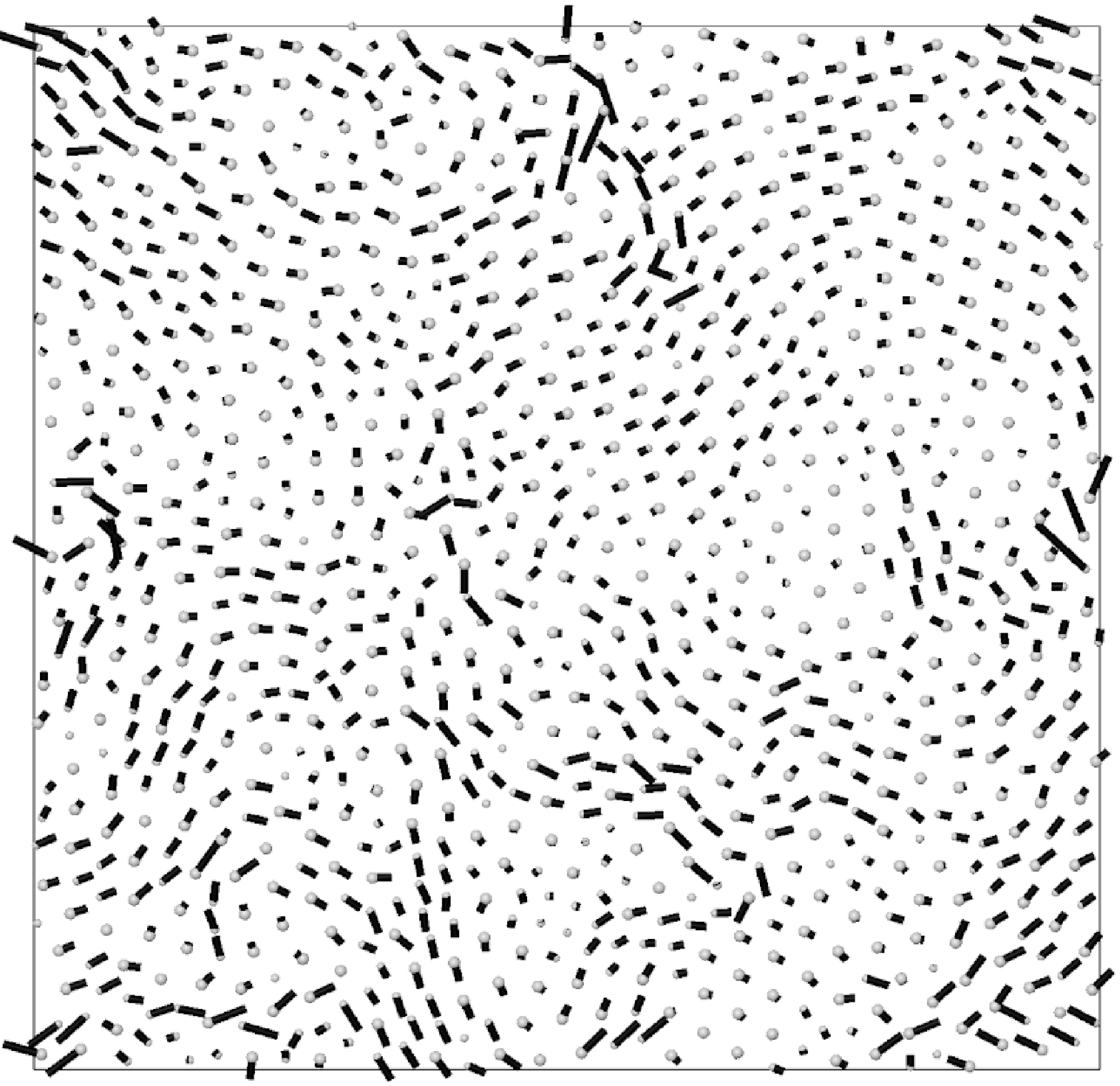}
  \caption{Lowest frequency eigenmodes for $N=1024$ bidisperse discs in $2D$,
    at two extreme values of $\phi - \phi_{c}$: $1\times 10^{-1}$
    (left), and $1\times 10^{-8}$ (right). The dots represent the
    centers of the particle and the black lines their polarization
    vectors.}
\label{fig4}
\end{figure}

The strong departure from the low-frequency Debye density of states
suggests that the eigenmodes are poorly characterized by simple plane
waves \cite{leo13}. We illustrate this point in Fig.~\ref{fig4} where
we show the lowest frequency modes for $2D$ harmonic systems at the
two extreme values of $(\phi - \phi_{c}) =1\times 10^{-1}$ and
$1\times 10^{-8}$. We show 2D results here for ease of visualization.
These correspond to modes below and above $\omega^{*}$, respectively,
for those packing fractions. In the compressed system at
$\omega<\omega^{*}$ the eigenmode retains identifiable correlated
plane-wave-like character, consistent with the more Debye-like
behavior of $\mathcal{D}(\omega)$ as $\omega \rightarrow 0$. In
contrast, when $\omega > \omega^{*}$ as in the system closest to
$\phi_{c}$, all plane-wave character is lost. This picture suggests
that $\xi_{T}$ represents the length scale above which one can average
over disorder: wavevectors of magnitude $k>2 \pi / \xi_{T}$ do not
have long enough wavelengths for effective averaging to take place.
This is related to the criterion that characterizes the
elastic-granular crossover in a Lennard-Jones glass
\cite{wittmer1}.

Our results for $\omega^{*}$ and $\xi_{T}$ motivated two recent
theoretical papers that treat different aspects of the
zero-temperature jamming transition. The first paper, by Wyart {\it et
  al.} \cite{wyart1}, shows that the constant density of states is a
direct consequence of the system being at what is called an isostatic
point, where the number of constraints precisely equals the number of
degrees of freedom \cite{ohern2,ohern3,moukarzel1}. This theory
predicts $\Omega=1/2$ and $\Omega=3/4$ for the harmonic and Hertzian
potentials, respectively, in excellent agreement with our simulations.
It also predicts a new length scale, $\ell \propto (\phi -
\phi_{c})^{-1/2}$.  This is not the same scaling as we have found for
$\xi_{T}$, but it is consistent with the scaling of $\xi_{L}$
predicted by the speed of sound argument (Eq.~\ref{equation3}).  The
length scale defined in Ref.~\onlinecite{wyart1} results from a
competition between bulk mechanical stability and boundary-induced
mode softening, and has not yet been explored by simulation.

A second theory by Schwarz {\it et al.} \cite{jen1} makes an analogy
between jamming at $\phi_{c}$ and the onset of $k$-core percolation.
The $k$-fold coordination required for each site corresponds to the
$D+1$-fold coordination required for a particle to be locally stable
in $D$ dimensions. For the Bethe lattice, Schwarz {\it et al.}
\cite{jen1} find exponents for the number of overlapping neighbors per
particle and the singular part of the shear modulus that are in good
agreement with our simulations \cite{ohern2}. Moreover, they find a
correlation length exponent of $\nu=1/4$, in agreement with
Eq.~\ref{equation2}.  Interestingly, the Bethe lattice calculations
\cite{jen1} also suggest the presence of another length scale with an
exponent of $\nu^{\#}=1/2$, in possible agreement with predictions of
Wyart {\it et al.} \cite{wyart1} and our argument for $\xi_{L}$.

O'Hern {\it et al.}  \cite{ohern2,ohern3} found another
correlation-length exponent by measuring the shift in the peak of the
distribution of jamming thresholds as the system size increases
\cite{ohern2,ohern3}. This exponent was measured to be $\tilde \nu
\approx 0.7$ in both 2 and 3 dimensions. Finally, yet another
diverging length scale has been measured in simulations by Drocco {\it
  et al.}  \cite{reichhardt1} as the jamming threshold is approached
from below. They measured the distance over which a slowly-moving
particle disturbs the surrounding packing and found $\nu_{-} =
0.6-0.7$, in good agreement with the finite-size scaling results of
O'Hern {\it et al.} \cite{ohern2,ohern3}.

There are at least three length scales that are important in
describing the jamming/unjamming transition: (i) The interparticle
overlap distance, which goes to zero at $\phi_{c}$ with a power-law
exponent of approximately 1.0 \cite{ohern3}. (ii) the length scale
determined from finite-size scaling \cite{ohern2,ohern3} as well as
the length \cite{reichhardt1} calculated as the transition is
approached from low densities. These diverge with an exponent of
roughly 0.7. (iii) the diverging length scale presented here on the
high-density side of the transition, determined from transverse
vibrations; this diverges with an exponent of 0.24. Theories
\cite{wyart1,jen1} and a scaling argument based on our data
(Eq.~\ref{equation3}) suggest a fourth length scale with exponent
$0.5$. It is still not clear how all these different length scales are
tied together. That different exponents are observed on different
sides of the transition suggests that the jammed phase may always be
separated from the unjammed one by a singularity.

Disordered or glassy systems characteristically show an excess in the
number of low-energy excitations.  In particular, there is a Boson
peak that can increase in magnitude and shift to lower frequency as
the glass transition is approached~\cite{buchenau4,buchenau5}.
However, nowhere are these two effects more clearly demonstrated than
at the onset of unjamming. There $\mathcal{D}(\omega)$ approaches a
constant as $\omega \rightarrow 0$ instead of dropping to zero,
implying, as shown in Fig.~\ref{fig1}(b), that the Boson peak diverges
and the peak position, $\omega^{*}$, shifts all the way to zero
frequency as the rigidity disappears. This is in accord with
random-network theories \cite{parisi1} and numerical studies imposing
disordered force constants on a lattice \cite{schirmacher1}.
The behavior we observe is clearly associated with the critical nature
of the transition and thus provides a natural way of describing these
effects in terms of a diverging length scale.  Our studies also
indicate that the enhancement of the low-frequency modes is purely a
geometrical phenomenon.

In conclusion, we have studied the properties of the jamming/unjamming
transition at zero temperature. Despite the discontinuity in the
number of interacting neighbors, we find that the loss of rigidity is
characterized by a diverging length scale.  Moreover, we note that the
jump in the number of interacting neighbors is universal, and given by
the isostatic condition. This indicates that unjamming may be
described as a second-order transition with a universal jump in the
order parameter.

\acknowledgments

We thank Lincoln Chayes, Gary Grest, Corey O'Hern, Jennifer Schwarz,
Thomas Witten, and Matthieu Wyart for insightful discussions. We
gratefully acknowledge the support of NSF-DMR-0087349 (AJL),
NSF-DMR-0352777 (SRN), DE-FG02-03ER46087 AJL,LES), and
DE-FG02-03ER46088 (SRN,LES).

\end{document}